\documentstyle[12pt]{article}

  \let\la=\lambda

\def\0{\over } \def\1{\vec }     \def\2{{1\over2}} \def\4{{1\over4}}
\def\5{\bar }  \def\6{\partial } \def\7#1{{#1}\llap{/}}
\def\8#1{{\textstyle{#1}}}       \def\9#1{{\bf {#1}}}
 \def\llp{\hbox to 0pt{\hss /\hskip1.5pt}}
\def\llo{\hbox to 0.2pt{\hss /}} \def\llq{\hbox to 0pt{\hss /\hskip0.5pt}}
\def\so{\supset\hbox to 0pt{\hss $\displaystyle -$\hskip1pt}}

\def\<{\langle } \def\>{\rangle }

\let\nn=\nonumber
\def\bea{\begin{eqnarray}} \def\eea{\end{eqnarray}}
\def\beann{\begin{eqnarray*}} \def\eeann{\end{eqnarray*}}
\def\beq{\begin{equation}} \def\eeq{\end{equation}}

\date{}
\title{
{\large\rm DESY 95-077}\hfill{\large\tt ISSN 0418-9833}\\
{\large\rm April 1995}\hfill\vspace*{3.5cm}\\
A Parton Model for Diffractive Processes\\
in Deep Inelastic Scattering}
\author{W. Buchm\"{u}ller and A. Hebecker\\
{\normalsize\it Deutsches Elektronen-Synchrotron DESY, 22603 Hamburg, Germany}
\vspace*{3.5cm}\\
}
\begin{document}

\setlength{\baselineskip}{18pt}
\maketitle
\begin{abstract}
\noindent
We demonstrate that the global properties of the ``rapidity gap'' events,
observed at HERA, can be understood based on electron-gluon scattering
as the underlying partonic process. Using the measured inclusive structure
function $F_2$ to determine the parameters of the parton model, the
diffractive structure function $F_2^D$ is predicted. The ratio of diffractive
and inclusive cross sections, $R_D = \sigma_D/\sigma_{incl}\simeq 1/9$, is
determined by the probability of the produced quark-antiquark pair to evolve
into a colour singlet state. This colour singlet cluster may fragment into
hadrons independently of the proton remnant, yielding the observed gap in
rapidity.
\end{abstract}
\thispagestyle{empty}
\newpage
In the ``rapidity gap'' events, observed in deep inelastic scattering
at HERA \cite{zeus}-\cite{h12},
the detected hadronic final state has small invariant mass,
and it is separated by a gap in rapidity from the proton beam
direction. The absence of a hadronic energy flow between proton
remnant and current fragment suggests that in the scattering process
a colour neutral part of the proton is stripped off which fragments
independently of the proton.  In analogy to hadronic processes
of similar kind the ``rapidity gap'' events are also called
``diffractive'' events.

The measured ``diffractive'' cross section is not suppressed at
large values of the momentum transfer $Q^2$ relative to the
inclusive cross section. Such a ``leading twist'' behaviour is usually
regarded as evidence for scattering on point-like constituents.
This interpretation, however, appears to be in conflict with the fact
that quarks and gluons, the constituents of the proton, carry colour.
Hence, one expects the formation of jets in the final state with a
hadronic energy flow familiar from ordinary deep inelastic scattering
events without a rapidity gap.

In the following we shall demonstrate that this puzzle can be resolved
by taking non-perturbative fragmentation effects into account.
We start from the production of a quark-antiquark pair in electron-gluon
scattering as basic partonic process.
Immediately after their production quark and antiquark
propagate in the colour field of the proton. With a certain probability,
approximated by a statistical weight factor, the
quark-antiquark pair evolves into a colour singlet parton cluster
which can escape from the proton and fragment independently into hadrons.
The cross section for diffractive events can then be calculated in terms
of the statistical weight factor, the quark-antiquark production cross section
and the gluon density which we determine in terms of the inclusive
structure function $F_2$. Our approach is related to previous
work on ``aligned jet models'' \cite{bj}-\cite{niza} and ``wee parton
lumps'' in deep inelastic scattering \cite{bu1,bu2}.

Quark-antiquark pairs are produced in electron-gluon scattering. The
relevant kinematic variables are (cf. fig. 1)
\beq
s=(P+k)^2\ ,\ Q^2=-q^2=xys\ ,\ x={Q^2\over 2 P\cdot q}\ ,
\eeq
which characterize inclusive deep inelastic scattering, and the invariant
mass $M$ of the quark-antiquark pair,
\beq
M^2=(q+p_g)^2\ .
\eeq
With $\vec{p}_g = \xi \vec{P}$ and $-p_g^2 = m_g^2 \ll Q^2,M^2$ one has,
\beq
\beta \equiv {Q^2\over Q^2+M^2} \simeq {x\over \xi}\ .
\eeq

The differential cross section for the inclusive production of
quark-antiquark pairs is given by \cite{field}
\beq\label{xep}
{d\sigma(ep\rightarrow e(q\bar{q})X)\over dx dQ^2 d\xi}
= {\alpha\over \pi x Q^2} g(\xi)\left(\left(1-y+{y^2\over 2}\right)
\left(\sigma_T+\sigma_L\right)-{y^2\over 2}\sigma_L\right)\ .
\eeq
Here $g(\xi)$ is the gluon density, and the cross section
$\sigma_{T(L)}$ is obtained by integrating the differential parton
cross section over the momentum transfer $t=(q-l')^2$ (cf. fig. 1)
from $t_{min}=-Q^2/\beta$ to $t_{max}=-m_g^2/\beta$,
\beq\label{xgag}
\sigma_{T(L)} = \int_{t_{min}}^{t_{max}}{d\sigma_{T(L)}\over dt}\ .
\eeq
The differential parton cross sections (see fig. 1 and the crossed process)
read \cite{field}
\bea
{d\sigma_T\over dt} &=&
{\pi\alpha\alpha_s \over Q^4}\sum_q e_q^2\beta^2
\left(4\beta(1-\beta)+{u\over t}+{t\over u}+{2Q^2\over tu}(t+u+Q^2)\right.
\nn \\
&&\left.\hspace{3cm} -Q^2 m_g^2\left({1\over u^2}+{1\over t^2}\right)\right),\\
{d\sigma_L\over dt} &=&
{8\pi\alpha\alpha_s \over Q^4}\sum_q e_q^2 \beta^3(1-\beta)\ ,
\eea
where the sum extends over all quarks whose mass is small compared to $Q$.

{}From eqs. (\ref{xep}) and (\ref{xgag}) one obtains the contribution to
the inclusive structure function $F_2(x,Q^2)$,
\beq\label{f2g}
\Delta^{(g)}F_2(x,Q^2)= x{\alpha_s\over 2\pi} \sum_q e_q^2
\int_x^1{d\xi\over \xi}g(\xi)
\left((\beta^2+(1-\beta)^2)\ln{Q^2\over m_g^2 \beta^2}
- 2 + 6\beta(1-\beta) \right)\ .
\eeq
This is the result in the ``massive gluon'' scheme \cite{field}. The virtuality
$-m_g^2$ of the gluon regularizes the collinear divergence at $t=0$. In the
more familiar minimal subtraction scheme one obtains a similar expression,
where $m_g$ is replaced by the subtraction scale  $\mu$,
and the finite part is different. At one-loop order,
the complete expression for the structure function $F_2$ reads,
\beq\label{f2}
F_2(x,Q^2) = x \sum_q e_q^2
\int_x^1{d\xi\over \xi}\left(S_q(\xi)+S_{\bar{q}}(\xi)\right)
\left(\delta(1-\beta)+{\cal O}(\alpha_s)\right) + \Delta^{(g)}F_2(x,Q^2)\ .
\eeq
Here $S_q$ and $S_{\bar{q}}$ are quark density and antiquark density,
respectively.

At small values of $\xi$ one has,
\beq\label{approx}
S_q(\xi)\ ,\ S_{\bar{q}}(\xi)\ \ll\ g(\xi)\ .
\eeq
For simplicity, we shall neglect in the following the quark contribution
to $F_2$. In this approximation we shall obtain a parameter free
prediction for the diffractive structure function. Note, that $S_q(\xi)$
is different from $S_q(\xi,Q^2)$, the quark density at scale $Q^2$.
Neglecting $S_q(\xi)$ in eq. (\ref{f2}) essentially amounts to
calculating $S_q(\xi,Q^2)$ in terms of an ``intrinsic'' gluon density
$g(\xi)$, corresponding to a scale $m_g = {\cal O}$(1 GeV$^2$). A more
complete analysis keeping the contribution from $S_q(\xi)$ would require
information about the longitudinal structure function $F_L$ in order
to predict the diffractive structure function.

For the gluon density at small values of $\xi$ we use the usual
parameterization,
\beq\label{gluon}
g(\xi)\ =\ A_g\ \xi^{-1-\lambda}\ ,
\eeq
where $A_g$ is a constant. Inserting eq. (\ref{gluon}) into eq. (\ref{f2g})
we can now evaluate the inclusive structure function $F_2$,
\beq\label{f2app}
F_2(x,Q^2)= {\alpha_s\over 2\pi} \sum_q e_q^2 x g(x)
\int_x^1 d\beta \beta^{\lambda}
\left((\beta^2+(1-\beta)^2)\ln{Q^2\over m_g^2 \beta^2}
- 2 + 6\beta(1-\beta) \right)\ .
\eeq
A simple approximation for $F_2$, valid at small $x$, is obtained by
choosing $x=0$ as lower limit of integration, and by setting $\lambda = 0$
in the integrand. This yields ($x \ll 1$),
\beq\label{fit}
F_2(x,Q^2) \simeq {\alpha_s\over 3\pi}\sum_q e_q^2 x g(x)
\left({2\over 3} + \ln{Q^2\over m_g^2}\right)\ .
\eeq
This expression provides a good description of the H1 measurement of the
structure function \cite{h13} for $\lambda=0.23$, $m_g=1.0$ GeV
and $A_g \alpha_s \sum_q e_q^2 = 0.61$. For
these parameters (\ref{fit}) is very close to the phenomenological fit
of the structure function $F_2$ given in \cite{h13}. We have also
evaluated the expression (\ref{f2app}) without any approximation.
Instead of the simple result (\ref{fit}) one then obtains some
cumbersome function of $x$ and $\lambda$. Comparison with the $H1$ data
yields again $\lambda=0.23$, whereas  the value of $m_g$ is now $0.8$ GeV.
The following discussion will be based on the simple analytical expression
(\ref{fit}).

We are now ready to calculate the diffractive structure
function. The main idea is that the quark-antiquark pair, originally
produced in a colour octet state,
changes its colour randomly through further soft
interactions with the colour field of the proton remnant.
Hence, the quark-antiquark pair evolves into a parton cluster which separates
from the proton remnant with some probability $P_8$ in a colour octet state,
and with probability $P_1 = 1 - P_8$ in a colour singlet state.
In the first case, there is a colour flow between proton remnant and current
fragment leading to the typical hadronic final state. In the latter case,
however, the colour singlet final state may fragment independently of the
proton remnant, yielding a gap in rapidity.
For a sufficiently fast rotation of the colour
spin of quark and antiquark, the probabilities
should simply be given by the statistical weight factor accounting
for the possible states of the quark-antiquark pair, i.e.,
\beq\label{stat}
P_1 \simeq {1\over 9}\ , \ P_8 \simeq {8\over 9} \ .
\eeq
Similar ideas concerning the rotation of quarks in colour space have
been discussed by Nachtmann and Reiter \cite{nacht} in connection with
QCD-vacuum effects on hadron-hadron scattering.

In analogy with the usual inclusive structure functions,
the diffractive structure functions are defined as,
\beq
{d\sigma_D \over dx dQ^2 d\xi} = {4\pi \alpha^2 \over x Q^4}
\left(\left(1 - y + {y^2\over 2}\right)F_2^D(x,Q^2,\xi)
- {y^2\over 2} F_L^D(x,Q^2,\xi)\right)\ .
\eeq
The structure function $F_2^D$ is easily obtained from eq. (\ref{f2g}).
With $x=\beta\xi$, and including
the statistical weight factor (\ref{stat}), one obtains
\beq\label{f2d}
F_2^D(x,Q^2,\xi)\simeq {1\over 9}{\alpha_s\over 2\pi} \sum_q e_q^2
g(\xi)\bar{F}_2^D(\beta,Q^2)\ ,
\eeq
where
\beq\label{fpom}
\bar{F}_2^D(\beta,Q^2) = \beta
\left((\beta^2+(1-\beta)^2)\ln{Q^2\over m_g^2 \beta^2}
- 2 + 6\beta(1-\beta) \right)\ .
\eeq
Since the gluon density $g(\xi)$ and the mass scale $m_g$ have been
determined by the fit to the inclusive structure function $F_2$,
the diffractive structure function is unambiguously predicted,
including its normalization.

Let us discuss the main properties of this result. An immediate consequence
is the prediction of
the ratio of diffractive and inclusive cross sections. From
eqs. (\ref{f2g})-(\ref{approx}) and (\ref{f2d}) one obtains,
\beq
R_D ={\int_x^1 d\xi F_2^D(x,Q^2,\xi)\over F_2(x,Q^2)} \simeq {1\over 9}\ .
\eeq
Note, that within the model described above, this ratio
directly measures the probability of forming a colour singlet
parton cluster in the scattering process.

The form of the diffractive structure function (\ref{f2d}) is identical with
expressions obtained based on the idea of a ``pomeron structure function''
\cite{ing1}-\cite{kwie}. The interpretation of the ingredients, however,
is rather different. The ``pomeron flux factor'' is replaced by the density
of gluons inside the proton, which factorizes. The ``pomeron structure
function'' for partons with momentum fraction $\beta$ inside the ``pomeron''
is identified as the differential distribution for the production of a
quark-antiquark pair with invariant mass $M^2=Q^2(1-\beta)/\beta$.

The function $\bar{F}_2^D(\beta,Q^2)$ is plotted in fig. 2 for three
different values of $Q^2$. For intermediate values of $\beta$ between
$0.2$ and $0.6$ this function is rather flat. Approximating
$\bar{F}_2^D(\beta,Q^2)$ in this interval by $\bar{F}_2^D(0.4,Q^2)$,
a comparison of eqs. (\ref{f2}) and (\ref{f2d}) yields the
scaling relation \cite{bu2}
\beq\label{scaling}
F_2^D(x,Q^2,\xi) \simeq {D\over \xi} F_2(\xi,Q^2)\ ,
\eeq
where $D\simeq 0.04$, independent of $Q^2$. Note, that the $\beta$-spectrum
shown in fig. 2 is rather sensitive to the infrared cutoff $m_g$ at small and
large values of $\beta$.

The scaling relation (\ref{scaling}) provides a rather accurate
description of recent measurements of the diffractive structure function
by the H1 collaboration. The experimental data are consistent with
\beq
F_2^D(x,Q^2,\xi)\ \propto\  \ln(Q^2)\ \xi^{-n}\ ,
\eeq
where $n=1.19 \pm 0.06 \pm 0.07$ \cite{h12}. This is in good agreement with our
results eqs. (\ref{f2d}),(\ref{fpom}) with $\la=0.23$.

Let us finally verify that the model described above predicts indeed
the appearance of rapidity gaps. For simplicity, we shall consider the
rapidity of the antiquark with momentum $l$ in the $\gamma^* p$-rest frame
(cf. fig. 1). It is related to other kinematical variables by
\beq
\eta = {1\over 2}\ln\left[\xi(1-\beta)\frac{u+m_g^2\beta}{t+m_g^2\beta}\right]
\, .
\eeq
{}From eq. (\ref{xep}) one obtains,
\bea
{d\sigma_D\over dydQ^2d\xi d\eta} &=& {dt\over d\eta}
{d\sigma_D\over dy dQ^2 d\xi dt} \nn\\
&=& {dt\over d\eta} {1\over 9}{\alpha\over \pi y Q^2}g(\xi)
\left(\left(1-y-{y^2\over 2}\right)\left({d\sigma_T\over dt}
+{d\sigma_L\over dt}\right) - {y^2\over 2}{d\sigma_L\over dt}\right)\ .
\eea
The total diffractive cross section for a maximum rapidity $\eta_{max}$
can now be obtained by integrating over a specified kinematic domain, where
the rapidity of the antiquark is larger than the rapidity of the quark. The
reverse configurations yield the same contribution, resulting in a factor of 2.
Hence, one obtains
\beq
{d\sigma_D\over d\eta_{max}} =2 \int_{Q_1^2}^{Q_2^2} dQ^2 \int_{y_1}^{y_2} dy
\int_x^{\xi_{max}}d\xi {d\sigma_D\over dy dQ^2 d\xi d\eta}\ ,
\eeq
where $\xi_{max}=\mbox{min}\{1,x+\exp 2\eta_{max}\}$. Using the kinematic
boundaries $0.03 < y < 0.7$ and $7.5$ GeV$^2 < Q^2 < 70$ GeV$^2$ \cite{h12} one
obtains the distribution shown in fig. 3. Note, that the approximation
$m_g^2\ll Q^2$ has been used throughout the calculation. Above $\eta_{max}
\sim 2$ the diffractive
cross section is clearly negligible. Assuming that the final state hadrons
are produced in the rapidity interval spanned by the quark-antiquark pair,
one obtains the observed rapidity gap.

In summary, we have demonstrated that electron-gluon scattering can
account for the global properties of the ``rapidity gap'' events observed
at HERA provided the following two hypotheses concerning the formation
of the final state are correct: First, the initially produced quark-antiquark
pair evolves with a probability, given by a statistical weight factor, into a
colour singlet parton cluster which fragments independently of the proton
remnant; second, the rapidity range of the diffractive hadronic final state
is essentially given by the rapidity interval spanned by the produced
quark-antiquark pair.

This simple picture appears to provide a rather accurate description of
the observed diffractive events, including the total rate, the
$\xi$-dependence and the $Q^2$-dependence. This agreement with
experimental data may appear fortuitous, since a number of theoretical
issues still remain to be settled. These include the non-perturbative formation
of the colour singlet cluster and the role of the infrared cutoff. Our results
indicate, that an appropriate starting point for the evaluation of
the inclusive as well as the diffractive structure functions at small
values of $x$ may be a semiclassical approach where ``wee partons''
are treated as a classical colour field. We expect that such a framework
will lead to results very similar to the ones described above.

We would like to thank G.~Ingelman, H.-G.~Kohrs, M.~Kr\"amer, M.~L\"uscher and
P.~Zerwas for valuable discussions.

\vspace{2cm}
\noindent
\\
{\bf\large Figure captions}\\
\\
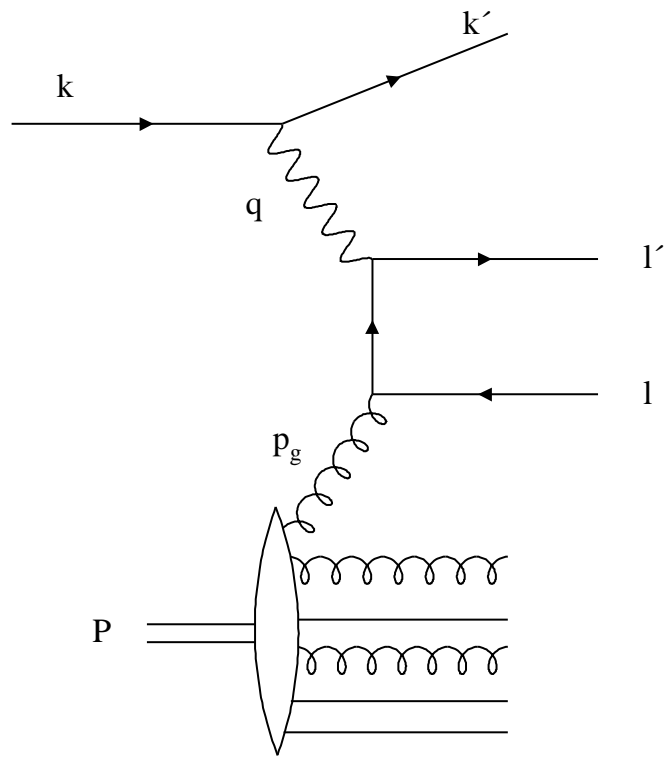
{\bf Fig.1} Quark-antiquark pair production in electron-gluon scattering.\\
\\
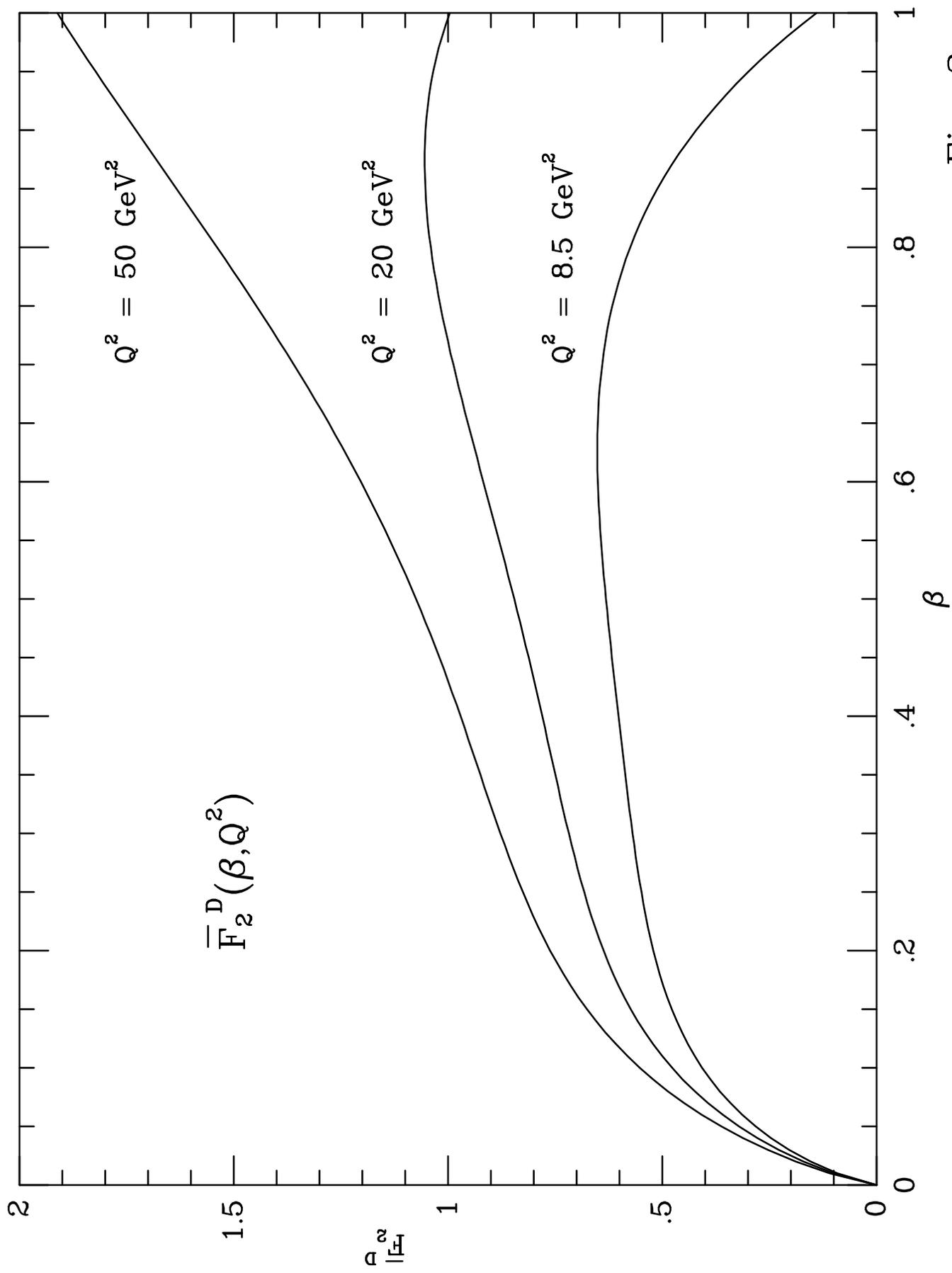
{\bf Fig.2} Dependence of the diffractive structure function on $\beta$ and
$Q^2$.\\
\\
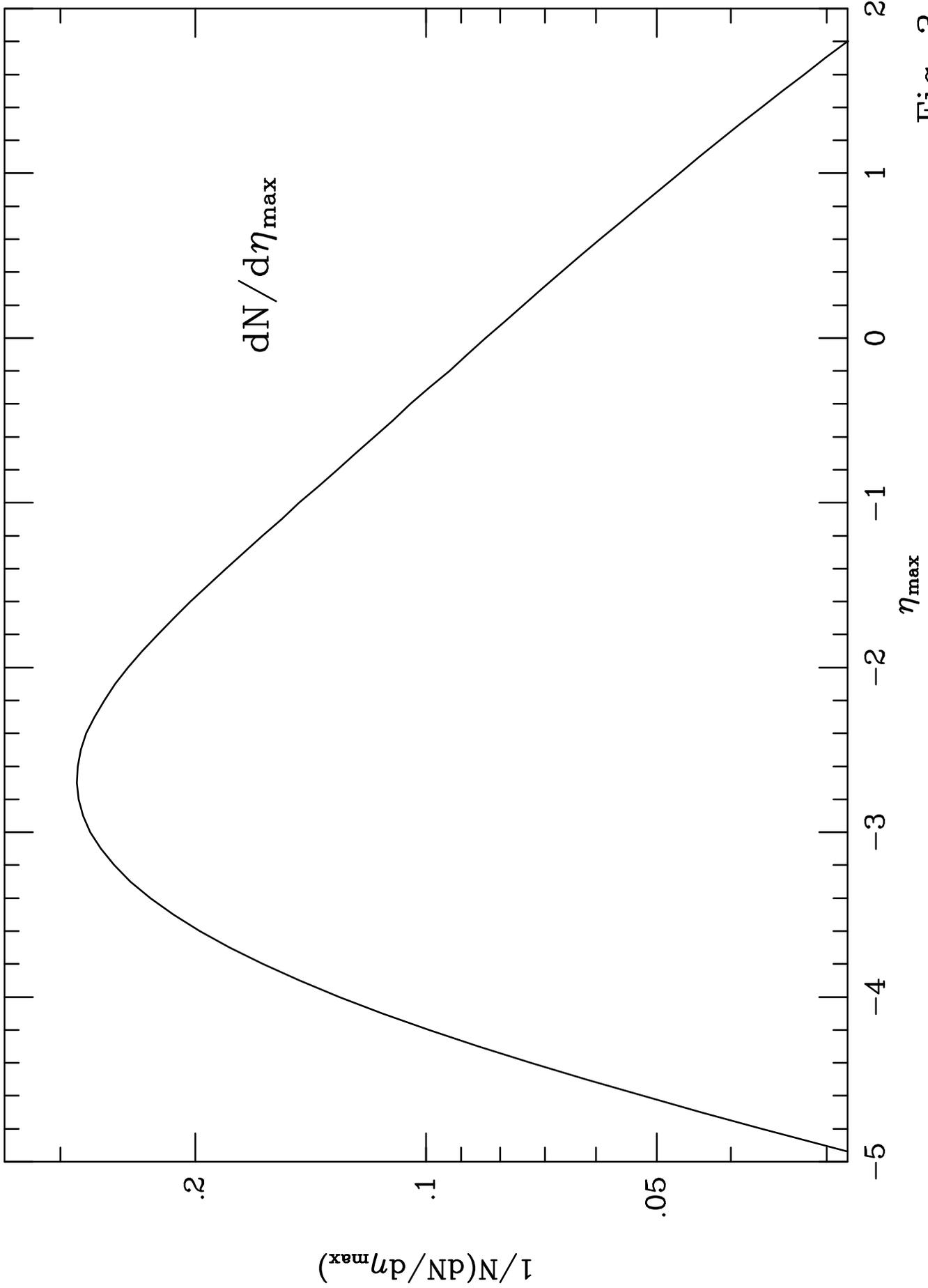
{\bf Fig.3} Distribution of the maximal rapidity in the $\gamma^*p$-rest frame
defined by the most forward quark or antiquark.\\
\\
\end{document}